# Dose Constraints for High-Resolution Imaging of Biological Specimens with Extreme Ultraviolet and Soft X-ray radiation

Chang Liu[a,b,c*], Leona Licht[a,b,c], and Jan Rothhardt [a,b,c,d]

*[a]Institute of Applied Physics, Abbe Center of Photonics, Friedrich-Schiller-Universität Jena, Albert-Einstein-Str. 15, 07745 Jena, Germany*
*[b]Helmholtz-Institute Jena, Fröbelstieg 3, 07743 Jena, Germany*
*[c]GSI Helmholtzzentrum für Schwerionenforschung, Planckstraße 1, 64291 Darmstadt, Germany*
*[d]Fraunhofer Institute for Applied Optics and Precision Engineering, Jena, Germany*

**Abstract**

We present a theoretical evaluation of radiation dose constraints for extreme ultraviolet (EUV) and soft X-ray microscopy. Our work particularly addresses the long-standing concern regarding strong absorption of EUV radiation in biological specimens. Using an established dose–resolution model, we compare hydrated and dehydrated cellular states and quantify the fluence required for nanoscale imaging. Our analysis identifies a *protein window* spanning photon energies from ~70 eV up to the carbon K-edge (284 eV), where EUV microscopy could in principle achieve sub-10 nm half-pitch resolution in dehydrated samples at doses well below the Henderson limit, thereby eliminating the need for cryogenic conditions. In this situation, the radiation dose required for EUV imaging is also substantially lower than what is required for comparable resolution in water window soft X-ray microscopy. Furthermore, EUV photons with sufficiently high energy exhibit penetration depths of µm-level in dehydrated biomatter, enabling exceptional amplitude and phase contrast through thin cellular regions and small cells. These findings provide quantitative guidelines for photon energy selection and establish the EUV *protein window* as a dose-efficient and physically viable modality for high-resolution, label-free, material-specific imaging of dehydrated biological matter.

[*] Corresponding author. Tel.: +49 03641 47810; e-mail: liu.chang@uni-jena.de.

## 1. Introduction

Soft X-rays (SXR) possess short wavelengths together with suitable absorption and scattering properties, making it ideally suited for high-resolution label-free imaging [1]. Pioneering work at synchrotron light sources established full-field transmission X-ray microscopy (TXM) [2] and scanning transmission X-ray microscopy (STXM) [3], which have evidenced major advancements and delivered sub-10 nm resolution in material science [4,5], enabled by advances of high-quality X-ray optics [6–8].

In particular, microscopes operating in the *water window* ( $E \approx 284 - 543\ eV$ ) have become a cornerstone for biological imaging [9–11]. The strong absorption difference between carbon and oxygen provides natural contrast between carbon-rich biomolecules (proteins, lipids, etc.) and the surrounding cytosol or nucleosol. State-of-the-art synchrotron-based TXMs in this regime now allow reliable label-free 3D visualization of whole hydrated cells up to 20 ~ 25 μm in diameter [12,13] with cryogenic preparation to mitigate radiation damage [14]. More recently, laboratory-scale water-window microscopes driven by compact laser–plasma sources have matured to provide routine access beyond large facilities [15], and have even entered the market as commercial products [16].

Complementary to lens-based approaches, X-ray diffraction imaging (XDM) has become an attractive bioimaging modality since its first demonstration on bacteria at Spring-8 [17]. By avoiding imaging optics, XDM intrinsically accesses both absorption and phase contrast, enabling an optimized contrast mechanism that improves dose efficiency compared to absorption-only imaging [18]. In this way, the achievable resolution is no longer constrained by lens fabrication but is only limited by the numerical aperture of the detector and ultimately, the signal-to-noise ratio (SNR) governed by the photon flux. Over the past two decades, developments at synchrotrons and free-electron lasers (FELs) have established XDM as a powerful tool for quantitative, label-free cellular imaging (Fig. 1). In particular, water window demonstrations have revealed ultra-structures in yeast [19], diatoms [20], mammalian cells [21] and fibroblast cells [22] in their native states, at resolutions down to ~ 30 nm. Beyond the water window, studies near the Fe L -edge (~ 750 eV) have enabled investigation of biological processes like biomineralization [23], while at hard X-rays (>5 keV), ptychographic X-ray computed tomography (PXCT) has achieved 3D reconstructions of thick tissues and organoids with sub-50 nm resolution [24–26].

Notably, sub-10 nm 2D/3D resolution has been repeatedly achieved in synchrotron-based TXM and XDM for non-organic materials [27–29], confirming that flux itself is not the limiting factor. Instead, radiation-induced damage fundamentally constrains biological resolution towards sub-10 nm in biological specimens, even under cryogenic conditions [1].

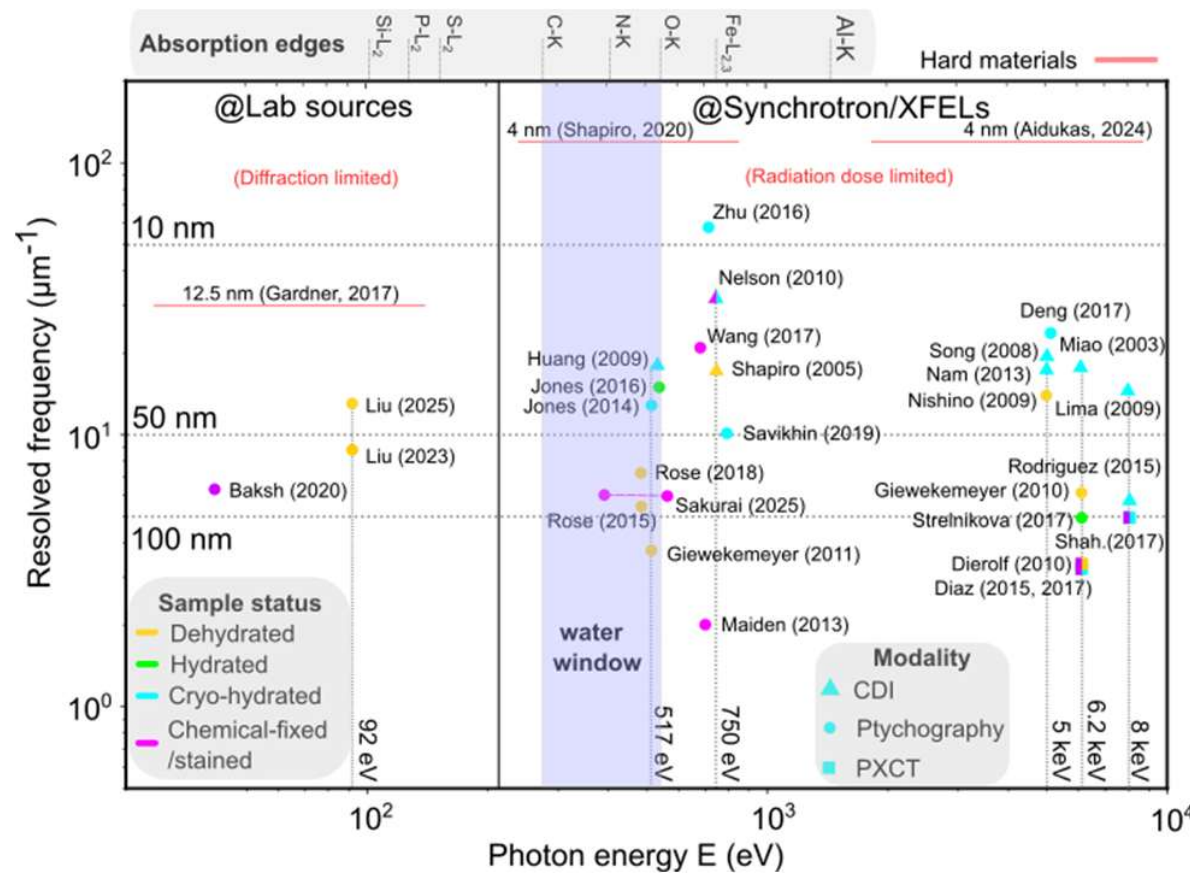

Fig. 1 Resolution versus photon energy in EUV and X-ray (< 10 keV) diffraction imaging demonstrations of biological specimens. Reported spatial resolutions for 2D and 3D diffraction imaging modalities, with color-coded by specimen preparation. The shaped region marks the water window. Non-biomaterial resolution benchmarks for EUV compact microscope, soft X-ray, and hard X-ray are marked, respectively.

Parallel to these developments, increasing attention of the microscopy community has turned towards the lower-energy end of the SXR spectrum, the extreme ultraviolet (EUV, $E \approx 12.4 - 124\ eV$). EUV coherent diffractive imaging (CDI), recognized as a nanoscale-resolution [30] and chemical sensitive [31] technique, has found broad use in nanotechnology [32], materials science [33–35], and wavefront sensing [36–38], whereas its exploration in biology has so far remained limited [39,40]. A prevailing view in the community is that the strong absorption of EUV photons in water renders hydrated state imaging impractical, and thus EUV has been overlooked.

However, a critical factor often missing in this discussion is the hydration state of biological samples. Under high-vacuum conditions typically intrinsic to EUV/ SXR diffraction imaging systems, cellular water content is rapidly lost in the absence of dedicated cryo-protection, yielding dehydrated samples while not necessarily preclude meaningful bioimaging. Certain specimens particularly benefit from dehydration. For example, bacterial cells are small and densely packed with biomolecules, drying such cells provides high contrast maps ideal for seeing storage granules and cell envelopes [40].

Previous studies show that dehydration of cells can be used as an alternative sample preparation to obtain ultrastructure information by X-ray microscopy, with chemical fixation to mitigate shrinkage artefacts [41]. Specifically, water loss increases the penetration depth of EUV radiation significantly, while contrast shifts from water-dominated absorption toward intrinsic atomic scattering phase contrast among principal biochemical elements C, O, N, P, S. These changes not only alleviate the strong water absorption but also provide exceptional stain-free chemical specificity in the EUV regime [42]. As such, EUV may find a unique niche in dehydrated state imaging. While not capturing living-state dynamics, dehydrated state imaging with enhanced contrast and simplified preparation offers a dose-efficient alternative to imaging in hydrated cells [43,44], particularly for high-throughput statistical studies, i.e., cell phenotyping [45]. Several XDM studies have employed air-dried or cryo-dried specimens [46,47], even within the water window [20,22], where hydrated state imaging is typically considered as the standard approach.

Pioneering theoretical work by Shen et al. [48] and Howell et al. [49] established the radiation dose-resolution scaling law in XDM based on the dose-fractionation theorem [50]. Notably, the Howell model, which incorporates the coherent phasor superposition and is highly relevant for EUV and soft X-ray XDM of biological specimens, predicted a 10 nm resolution limit for frozen hydrated cells using a protein-water voxel model. Following that, the required fluences and doses to image a feature in XDM have been calculated under different assumptions using this model [51–53]. In particular, Nave performed similar calculations for cellular components in cytosol, comparing pure absorption contrast with phase contrast [54], and further claimed that the required dose for achieving a given resolution is independent of the object size [55].

However, these analyses neither directly apply to EUV nor to dehydrated biological matter, where the dominant contrast mechanisms differ significantly. This leaves open questions in understanding the complementary potential of dehydrated-state imaging in EUV compared to cryo-hydrated approaches in the water window.

In this work, we provide a theoretical analysis of radiation dose requirements for XDM of dehydrated biological specimens, based on the dose-resolution scaling law [49]. We introduce the concept of a *protein window* spanning photon energies from ~ 70 eV up to the carbon K edge (284 eV), defined by the attenuation properties of proteins as the dominant biomolecular components in cells. In this spectral window, our analysis demonstrates that sub-10 nm resolution is theoretically achievable for dehydrated cells at doses well below the Henderson limit, where proteins are observed to begin fading out [56]. These findings directly address the longstanding concerns regarding EUV absorption and establish conditions under which EUV can serve as a dose-efficient and physically viable modality for high-resolution, stain-free, and chemical-sensitive biological imaging without the need for cryogenic cooling.

The simulation results (Section 3) are presented in the following structure: First, we analysed the EUV/SXR attenuation properties of major cellular components and cytosolic backgrounds, a step crucial for defining the protein window. Second, we determine the complex contrast and corresponding dose requirements for XDM of a hydrated cell model. Finally, we extend this investigation to the dehydrated cells, thereby highlighting the *protein window* as a dose-efficient energy regime for high-resolution biological imaging.

## 2. Theoretical Overview

In imaging systems, photon arrival at the detector follows Poisson statistics, such that the SNR scale with the square root of the detected photon number [1]

$$SNR = \sqrt{\bar{n}} \cdot \Theta \tag{1}$$

where $\bar{n}$ is the expected number of incident photons per voxel and $\Theta$ denotes the intrinsic contrast parameter between feature and background [57]. To reliably resolve a feature, the SNR must satisfy the widely-adopted Rose criterion [58]. This criterion, though derived in the context of direct imaging, can be extended to diffraction-based modalities, as the statistical requirement for distinguishing a feature against noise remains identical [1,49]. Hence, improving image fidelity requires either increased photon flux or stronger intrinsic contrast, with the latter being related to the optical properties of feature and background. These calculations assume that 2D and 3D reconstructions require comparable exposures when significance and resolution are matched, proven by Hegerl & Hoppe [50].

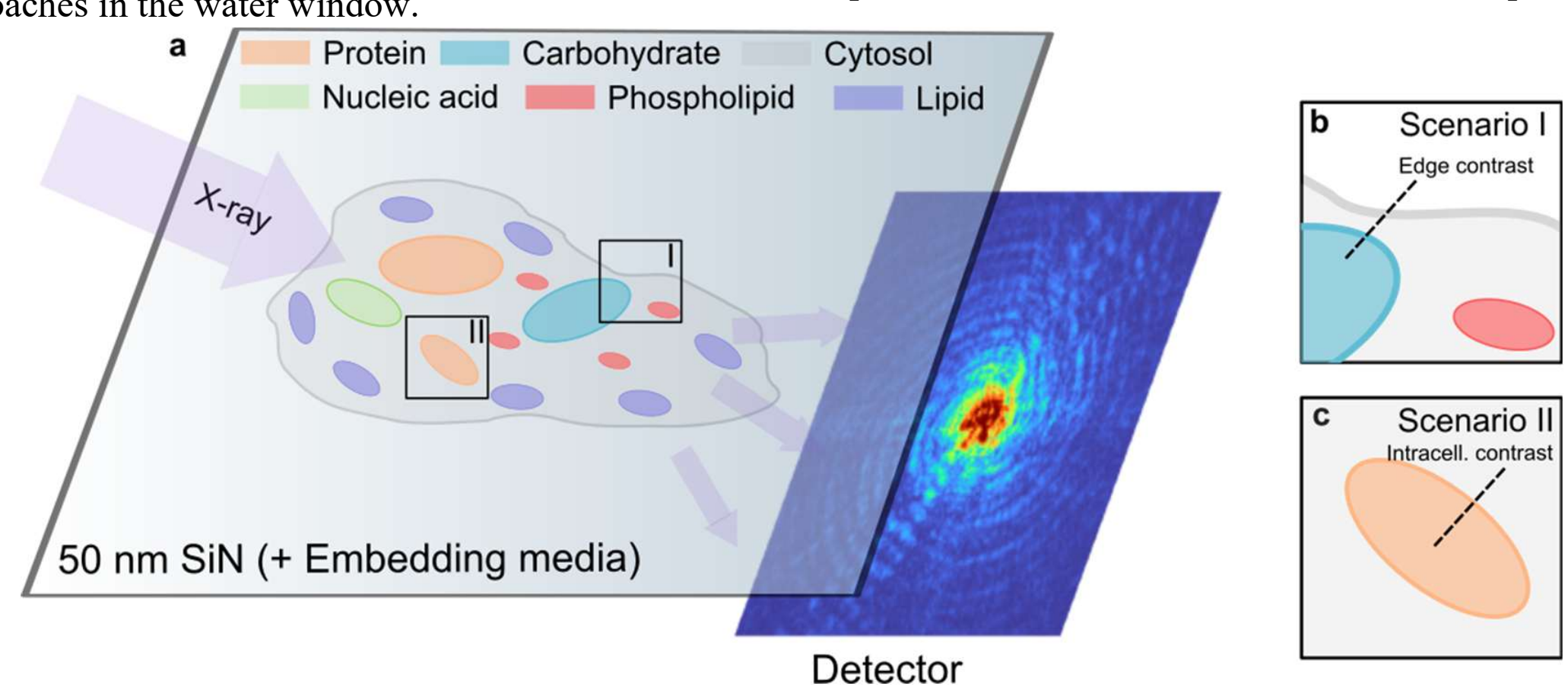

Fig. 2 Schematic of X-ray (including EUV) diffraction imaging model for biological specimens. a. cells supported on a 50 nm SiN membrane with optional embedding medium, leading to two contrast scenarios. b. Contrast scenario I: edge contrast between biomolecules and surrounding background. c. Contrast scenario II: intracellular contrast between biomolecules in cytoplasmic solutions with a baseline density.

## *2.1. Radiation fluence, dose and resolution scaling*

In the framework established by Howell et al. [49], the radiation dose can be related to the feature voxel size to be resolved. Considering a single voxel of dimensions $t^3$, the fluence $\mathcal{F}$ (incident photon number per unit area) is $\mathcal{F} = \bar{n}/t^2$. A fraction of these photons $\bar{n}_s = \mathcal{F}\sigma_s$ are coherently scattered, where $\sigma_s$ is the X-ray coherent scattering cross-section [9]. The energy deposition on the feature is quantified by the radiation dose [Gy]

$$\mathcal{D} = \frac{\mathcal{F}\mu h\nu}{\rho} = \frac{\bar{n}_s \mu h\nu}{\rho\sigma_s} \quad (2)$$

where $h\nu$ corresponding to photon energy, $\mu$ is the linear absorption coefficient (LAC), and $\rho$ is the mass density. By relating $\sigma_s$ to material properties, Howell showed that

$$\sigma_s = r_e^2\lambda^2|\tilde{\epsilon}_r|^2 t^4 \quad (3)$$

with $r_e$ the classical electron radius, $\lambda$ the illumination wavelength, and $\tilde{\epsilon}_r$ the relative complex electron density, directly related to the complex relative refractive index difference between a feature and its background (subscript 0) as

$$\tilde{\epsilon}_r = \frac{2\pi}{\lambda^2 r_e}\{(\delta - \delta_0) + i(\beta - \beta_0)\} \quad (4)$$

where $\delta$ and $\beta$ represent the real and imaginary parts of the refractive index, respectively. Substituting Eq. (3) into (2) yields the required dose

$$\mathcal{D}_t = \frac{\mu h\nu}{\rho} \cdot \mathcal{F}_t = \frac{\mu h\nu}{\rho} \cdot \frac{\bar{n}_s}{r_e^2\lambda^2|\tilde{\epsilon}_r|^2 t^4} \quad (5)$$

It should be noted that Eq. (5) assumes a single feature voxel under uniform illumination. For finite-thickness biological specimens, however, the fluence and dose have to be corrected by the overall transmission of the specimens [55], described by the Beer-lambert law as $T = e^{-\mu r}$, where $r$ is the path length through the specimen. Accordingly, the required incident fluence and dose are increased by a factor of $1/T$. This correction does not affect the intrinsic dose-resolution scaling and is negligible for thin, dehydrated specimens with a thickness below the penetration depth of radiation, while significant for hydrated cells with up to ~10 μm thickness.

The calculation of the required fluence and dose is further dependent on the Rose criterion SNR value $K$ adopted. $K = 5$ has been widely used [59] to define the fluence $\mathcal{F}_t$ and dose $D_t$ for a half-pitch resolution $t$ in this model. This is often interpreted as a practical threshold of $\bar{n}_s = 25$ scattered photons per voxel to ensure sufficient SNR [49]. Other analyses, however, have argued that because phase retrieval in CDI operates on amplitude rather than intensity, the effective photon requirement may be reduced to $\bar{n}_s = 6.25$ rather than 25 [60]. Moreover, in CDI the requirement has been calculated using $K = 3$ with field-of-view correction [52].

In this work, we assume that each voxel can be sufficiently illuminated (no attenuation correction) and adopt the conservative 25-photon threshold to enable consistent comparison with previous studies.

## *2.2. X-ray imaging contrast in biological cells*

According to Eq. (4), a feature can be imaged against its surrounding materials through both absorption and phase contrast. Predicting contrast in biological samples requires knowledge of the relative complex refractive index $(\Delta\delta, \Delta\beta)$ between feature and background. Away from absorption edges, phase contrast is dominated by $\Delta\delta$, whereas absorption contrast is determined by $\Delta\beta$. Following Howell et al. [49], we quantify contrast using the relative electron density $\tilde{\epsilon}_r$ (Eq. (4)), which enters the required fluence and dose expressions (Eq. (5)) with the optimized phase contrast, exploiting absorption and phase shift [18], and is most relevant to diffraction imaging. For clarity, we rename it as *complex contrast* throughout the manuscript. It should be noted that Nave evaluated the dose requirement for absorption-based TXM, where contrast is governed solely by $\Delta\beta$, proving the dose-efficiency within the water window [54].

Here we adopt a conceptual model of the cell in X-ray imaging (Fig. 2**a**), with specimens deposited on a thin, X-ray transparent membrane (e.g., 50 nm SiN), optionally embedded in media such as Araldite M and paraffin, for better preserving morphology. Two contrast scenarios arise based on model assumptions. First, edge contrast at the interface between cells and the surrounding medium (Fig. 2**b**). Second, intracellular contrast focuses on distinguishing specific biomolecular domains within the cytosolic or embedding medium (Fig. 2**c**).

Nave has treated the cellular interior as single effective objects (e.g., heterochromatin, lipid droplet, and mitochondrial membranes), concerning that current resolutions are insufficient to separate biomolecules in complex ultra-structures [54]. In contrast, we model the interior as a heterogeneous mixture of biomolecular classes (proteins, nucleic acids, carbohydrates, lipids, phospholipid) in cytosol, enabling explicit estimation of biomolecular contrast.

Atomic compositions and densities required for refractive indices estimation are obtained from the CXRO database [42] (Tabel S1 in supplementary material). Typical biomolecules have well-characterized atomic compositions taken from literatures [54,61]. For complex mixtures such as cytosol, effective atomic ratios and densities are derived from reported mass fractions [54], benchmarked by experimentally measured cellular density values [24]. The parameters for all modelled backgrounds (Table S2 in supplementary material) are subsequently used to calculate the fluence and dose for target resolutions via Eq. (5).

### *2.3. Dehydrated cell model*

Biological cells are predominantly composed of water ( 60% - 80% of total mass), with the remaining dry matter (protein, lipid, etc.) accounting for an overall cell density ~ 1.1 g/cm$^3$ [62]. Under high-vacuum conditions without cryogenic preservation, cellular water is rapidly lost, and specimens are effectively imaged in a dehydrated state. Accurate modelling of such conditions is therefore required for contrast and dose analysis.

Collectively, previous reports from electron microscopy (EM) [63], X-ray [41,64] and EUV [39] imaging studies indicate that properly dehydrated biological samples (both chemically-fixed and unfixed) undergo minimum shrinkage with negligible volume loss and geometry change. In alignment with these findings, we adopt a volume-invariant approximation to model a dehydrated cell, in which the overall geometry of the cell is preserved while the water mass is removed. For instance, based on reported mass fractions of the cytosolic mixture [54], we calculate the effective composition and density of a 'dehydrated cytosol' background (Table S2). This approximation provides a tractable and well-supported basis for estimating contrast and dose in EUV&X-ray imaging of dehydrated cells, while acknowledging that localized shrinkage may occur in practice.

## 3. Results

### *3.1. X-ray attenuation in biological cells*

Before analysing contrast and dose, it is essential to first examine the penetration length of X-rays in cells, since it sets the practical spectral windows for both TXM and XDM. Fig. 3**a** shows the calculated attenuation lengths of whole-cell and background models, including hydrated and dehydrated cytosol, embedding media, and water. Fig. 3**b** presents the attenuation lengths of representative biomolecules calculated from tabulated atomic compositions and densities [54]. Given that composition is highly variable across different cells, the analysis presented here adopts a representative, literature-derived mass composition model for a generic hydrated cell. Specifically, this model utilizes the fractional ratios of 70% water, 8% nucleic acid, 16% protein, 2% lipid, 3% carbohydrate, and 1% inorganic ions, as detailed in Ref. [65].

In the well-known water window (284 – 543 eV), water (cyan dash line) is relatively transparent while biomolecules absorb strongly (typical attenuation length < 1 µm, Fig. 3**b**), yielding strong absorption contrast that underpins the success of cryo-hydrated TXM/STXM. However, this strong absorption of biomolecules also restricts the specimen thickness, e.g., at 517 eV, 2 µm for a hydrated cell [54,65] (black line in Fig. 3**a**). The achievable resolutions of absorption-based TXM in hydrated cells rarely

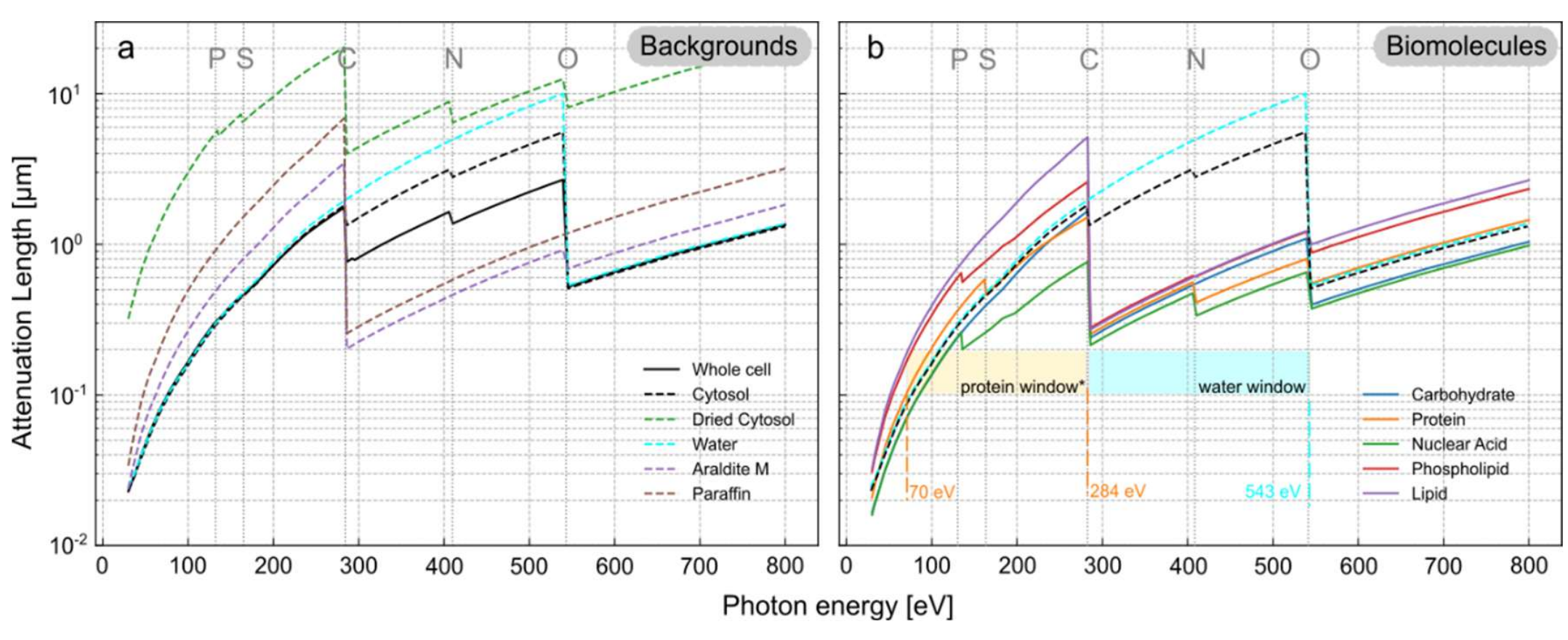

Fig. 3 a. Attenuation length of whole-cell and background models, including hydrated and dehydrated cytosol, embedding media (Araldite M, paraffin), and water. b. Attenuation length of representative biomolecules (protein, nucleic acid, carbohydrate, lipid, phospholipid) calculated from tabulated atomic compositions and densities. The *water window* (284–543 eV) is highlighted as the favourable regime for hydrated imaging due to the strong absorption difference between carbon and oxygen. Analogously, we introduce a *protein window** (≈70–284 eV), defined as the regime where the attenuation length of protein exceeds 100 nm, providing optimal conditions for dehydrated imaging. The whole-cell curve is derived from the weighted composition of major biomolecular classes and water (70% of mass), with detailed composition and density values summarized in Table S2.

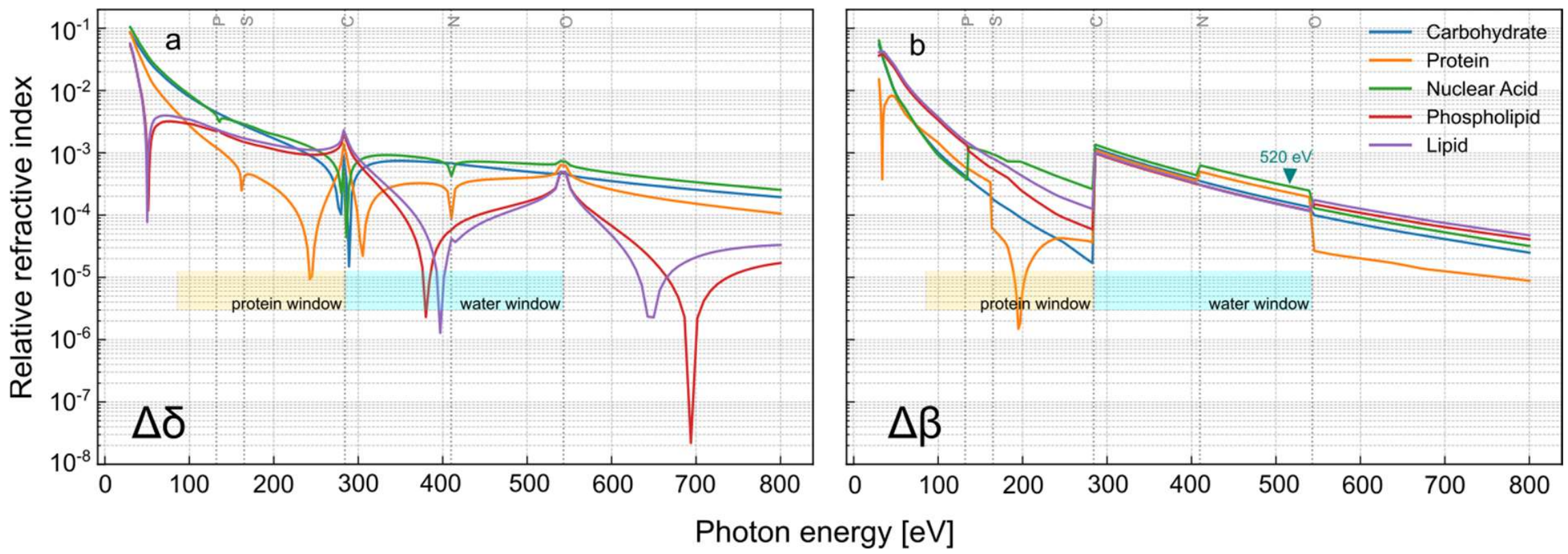

Fig. 4 Refractive (a, $\Delta\delta$) and absorptive (b, $\Delta\beta$) contrast of major biomolecular classes (protein, nucleic acid, carbohydrate, phospholipid, lipid) relative to cytosol under hydrated conditions.

surpass 30 nm even under cryogenic protection [11]. Moreover, the absorption contrast between different biomolecules in this regime remains modest (e.g., carbohydrate vs. protein, Fig. 3**b**).

Below the carbon K edge (284 eV), poor contrast and limited penetration depth in hydrated cells are expected. Nevertheless, biomolecules display major differences in attenuation (solid lines in Fig. 3**b**). In the spectral range from about 70 eV up to 284 eV, proteins, which represent the dominant fraction of dry cell mass, exhibit attenuation lengths exceeding 100 nm. We define this favourable regime as the *protein window* (70 – 284 eV), which provides optimal conditions for dehydrated state imaging. Especially at the high-energy end of this window, carbon-based biomolecules exhibit attenuation lengths of more than 1 µm, an order of magnitude longer than in the water window. Note that, previous research suggested a narrow spectral region immediately below the carbon K edge as so-called *carbon window* (247 - 283 eV) for dehydrated-state imaging [66]. Most importantly, the dehydrate cytosol (green dashed line in Fig. 3**a**), serving as a model of intracellular background after dehydration, reaches attenuation lengths up to 10 µm. In the protein window, the attenuation length of dehydrated cytosol significantly surpasses those of the major biomolecules. It thus remains comparatively transparent and creates a robust absorption contrast mechanism that is critical for achieving high-fidelity dehydrated-state imaging.

Furthermore, typical embedding media, such as Araldite M and paraffin (purple and brown dashed line in Fig. 3a), exhibit strong absorption across the water window (0.2 µm – 1 µm) and are transparent at the high-energy end of the protein window (> 1 µm).

Collectively, the presented attenuation analysis highlights that the protein window (70–284 eV) offers sufficient penetration depth and significant absorption contrast for dehydrated-state imaging. This specific regime offers a biochemical basis for stain-free, dose-efficient nanoscale microscopy, potentially even for thin tissue sections embedded in common media like Araldite or paraffin.

### *3.2. Hydrated-state imaging*

Hydrated-state imaging is considered first, as it represents a well-explored regime of TXM, providing a benchmark for understanding the influence of background medium. The intrinsic contrast of biomolecules relative to a hydrated cytosolic background was analysed in several pioneer works [49,54,55] and is used as a benchmark here.

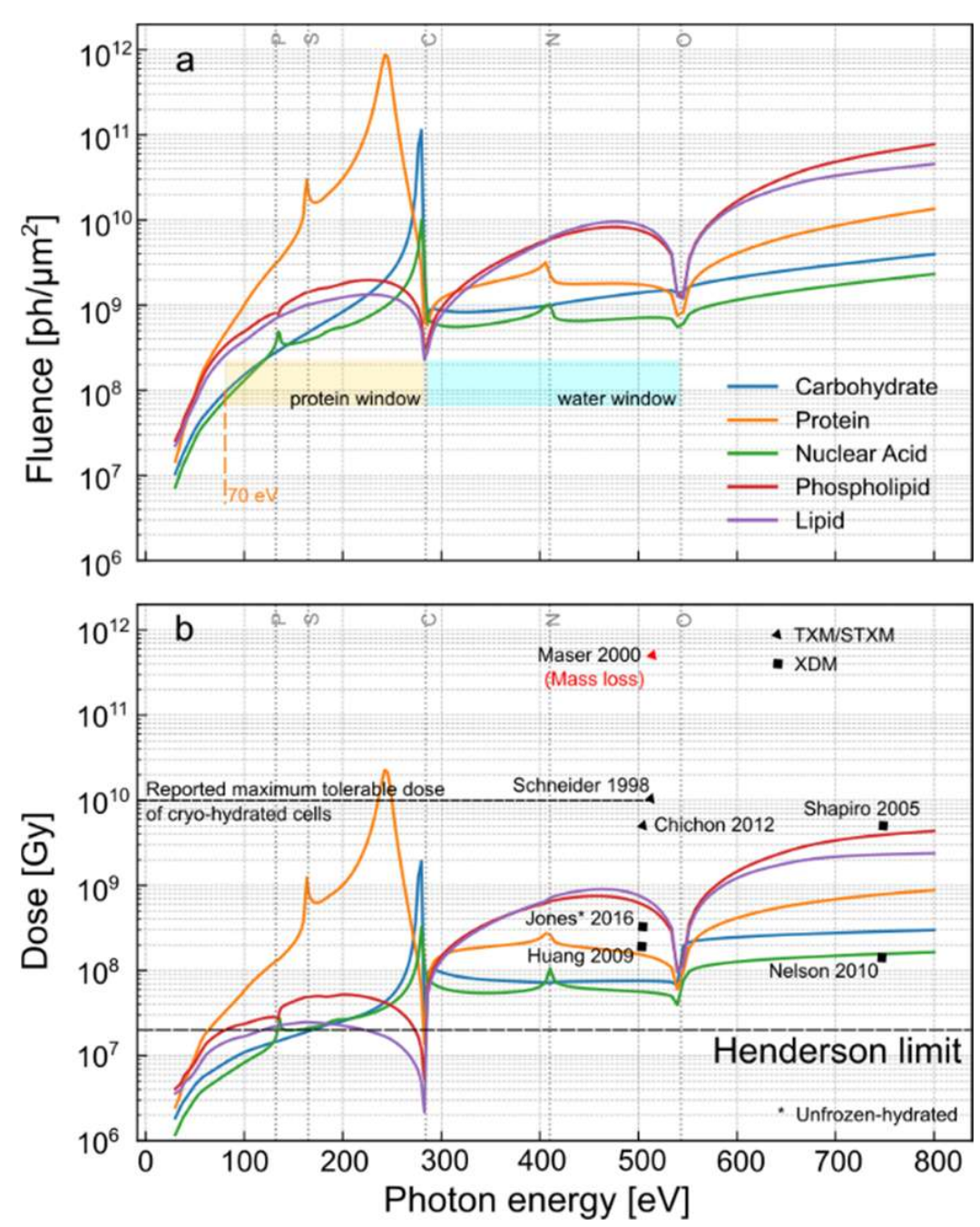

Fig. 5 Fluence (a) and dose (b) requirements for resolving 10 nm half-pitch biological molecules in cytosolic background under Rose criterion (K = 5). Henderson limit (black dashed line) is marked as general damage threshold. Reported dose values from SXR hydrated cell imaging experiments (including XDM, with claimed resolutions of 10 ~ 60 nm) are also plotted in (b).

Fig. 4 shows the refractive ($\Delta\delta$) and absorptive ($\Delta\beta$) differences across the SXR regime. Distinct dips in Fig. 4**a** and **b** appear where refractive index and absorptive coefficients of specific biomolecules intersect with that of cytosol, respectively. Overall, proteins exhibit the weakest contrast in both phase and absorption.

The corresponding fluence and dose for resolving 10 nm half-pitch features with complex contrast under the Rose criterion is shown in Fig. 5. The calculated fluence and dose for proteins are in good quantitative agreement with previous theoretical studies [49,54], supporting ~10 nm as a practical resolution limit for hydrated XDM. For comparison, dose estimates based solely on absorption contrast were additionally evaluated for TXM (see Supplementary Material) and benchmarked against prior work [54], confirming the relative dose advantages of diffraction imaging over lens-based approaches.

Across the full spectral range, the required dose frequently exceeds the Henderson dose limit (~$2 \times 10^7$ Gy), implying severe damage to molecular bonds. In the water window, dose requirements for resolving carbohydrates and nucleic acid remain below $10^8$ Gy, which is experimentally supported to be easily managed with cryogenic temperature [18,19,67] or by chemical fixation at room temperature [68], as indicated in Fig. 5**b**. Notably, STXM experiment with a dose of $5 \times 10^{11}$ Gy have been reported to induce mass loss even with cryo-cooling [69].

In the protein window, the applied model outputs a lower required dose (~$10^7$ Gy) except for proteins ( $10^{10}$ Gy). However, due to the strong water absorption (< 1 μm), hydrated-state imaging is impractical. At higher photon energies, above the oxygen K edge (543 eV), the water absorption again dominates, leading to higher dose requirements and necessitating cryogenic cooling [47,70].

Consequently, the spectral range between the nitrogen and oxygen edges remains particularly favourable, explaining why most TXM has been performed at the energies near 520 eV [13].

*3.3. Dehydrated-state imaging*

Next, calculations were performed relative to dehydrated cytosol as intracellular background. Fig. 6 shows the refractive ($\Delta\delta$) and absorptive ($\Delta\beta$) difference. In hydrated cells, water dominates absorption and reduces phase contrast within the water window, whereas dehydration removes water contribution and replaces it with vacuum, substantially enhancing complex contrast and increasing attenuation lengths by up to an order of magnitude (1~10 μm compared to < 1 μm in hydrated conditions).

Both $\Delta\delta$ and $\Delta\beta$ decrease with increasing photon energy, but distinct features emerge near elemental edges. Phase contrast (Fig. 6a) dominates below the carbon K-edge with characteristic dips appearing across carbon-based biomolecules. After the phosphorus (132 eV) and sulfur (165 eV) edges, biomolecules exhibit distinct attenuation, providing excellent intracellular absorption contrast in dehydrated samples.

Fig. 7 shows the required fluence and dose for resolving 10 nm half-pitch features under the Rose criterion, together with reported dose values from XDM experiments on dehydrated biological specimens. A comparison of Fig. 7b with Fig. 5b clearly demonstrates that eliminating the water from the specimen significantly enhances the radiation dose tolerance [71]. Most importantly, the removal of water also lowers the dose to achieve high-resolution contrast. Our analysis confirms that detecting a biomolecular voxel in a dehydrated environment demands substantially less dose than in hydrated cells, thereby enabling high-resolution imaging without the need for cryo-cooling [21,22,39].

The most significant effect appears in the dose requirements (Fig. 7b) below the carbon K-edge, i.e., protein window. For 10 nm half-pitch features, dehydration reduces the required dose by up to two orders of magnitude for proteins and falls below the Henderson limit. This dose demand reduction highlights the unique advantage that sub-10 nm resolution dehydrated XDM using lower photon energies becomes physically viable without cryogenic cooling of the sample.

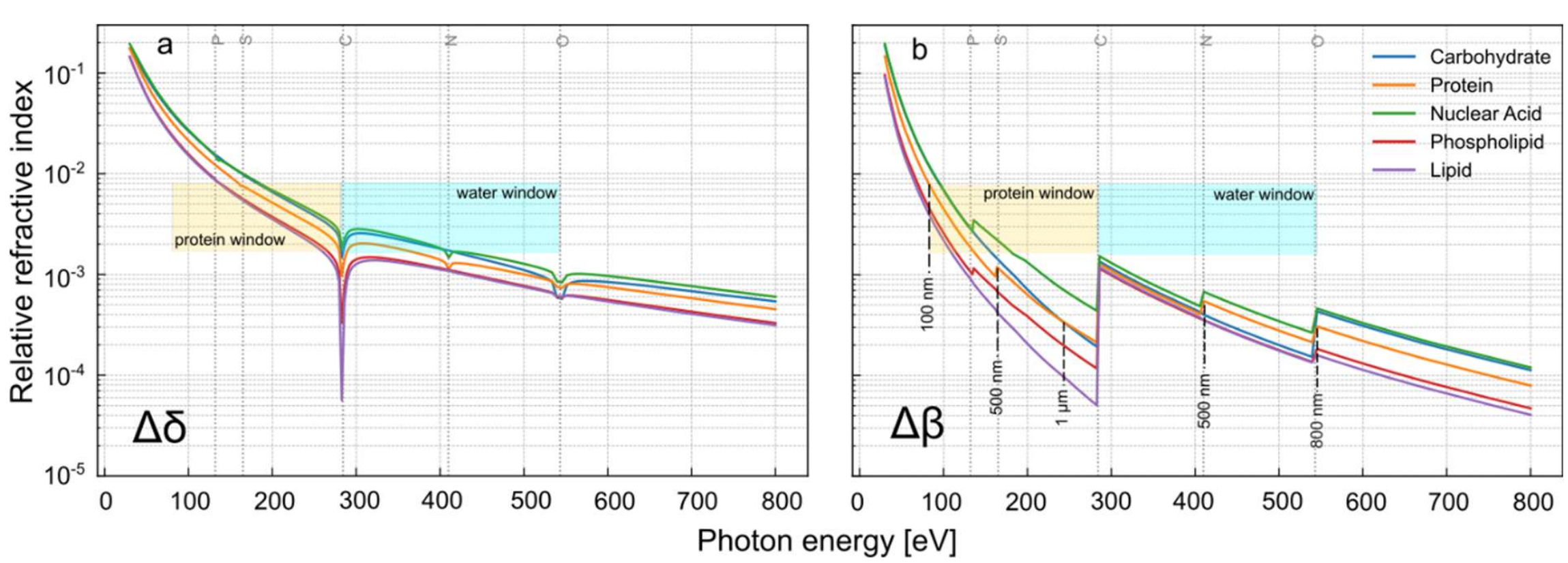

Fig. 6 Refractive (a, $\Delta\delta$) and absorptive (b, $\Delta\beta$) contrast of major biomolecular classes (protein, nucleic acid, carbohydrate, phospholipid, lipid) relative to cytosol under dehydrated conditions. The attenuation lengths of protein are marked in b as reference.

For completeness, it's worth noting that above the oxygen K-edge (543 eV), the required dose shows weak dependence on energy, as the $\lambda^{-2}$ scaling of the fluence is counterbalanced by the $\lambda$-dependent absorption coefficient $\mu$ in Eq. (5).

For embedded dehydrated tissue slices (e.g., paraffin-embedded sections), the absorption and scattering properties of embedding media resemble those of certain biomolecules, thereby reducing intrinsic contrast and increasing the dose requirement. Nevertheless, the required dose remains about an order of magnitude lower compared to fully hydrated specimens. An example is provided for tissues embedded in paraffin, summarized in the Supplementary material.

To conclude, these results identified imaging of dehydrated biological samples at photon energies in protein window (70 - 284 eV) as a dose-efficient regime, offering strong intrinsic contrast and capability of sub-10 nm cellular imaging with cryo-free sample preparation.

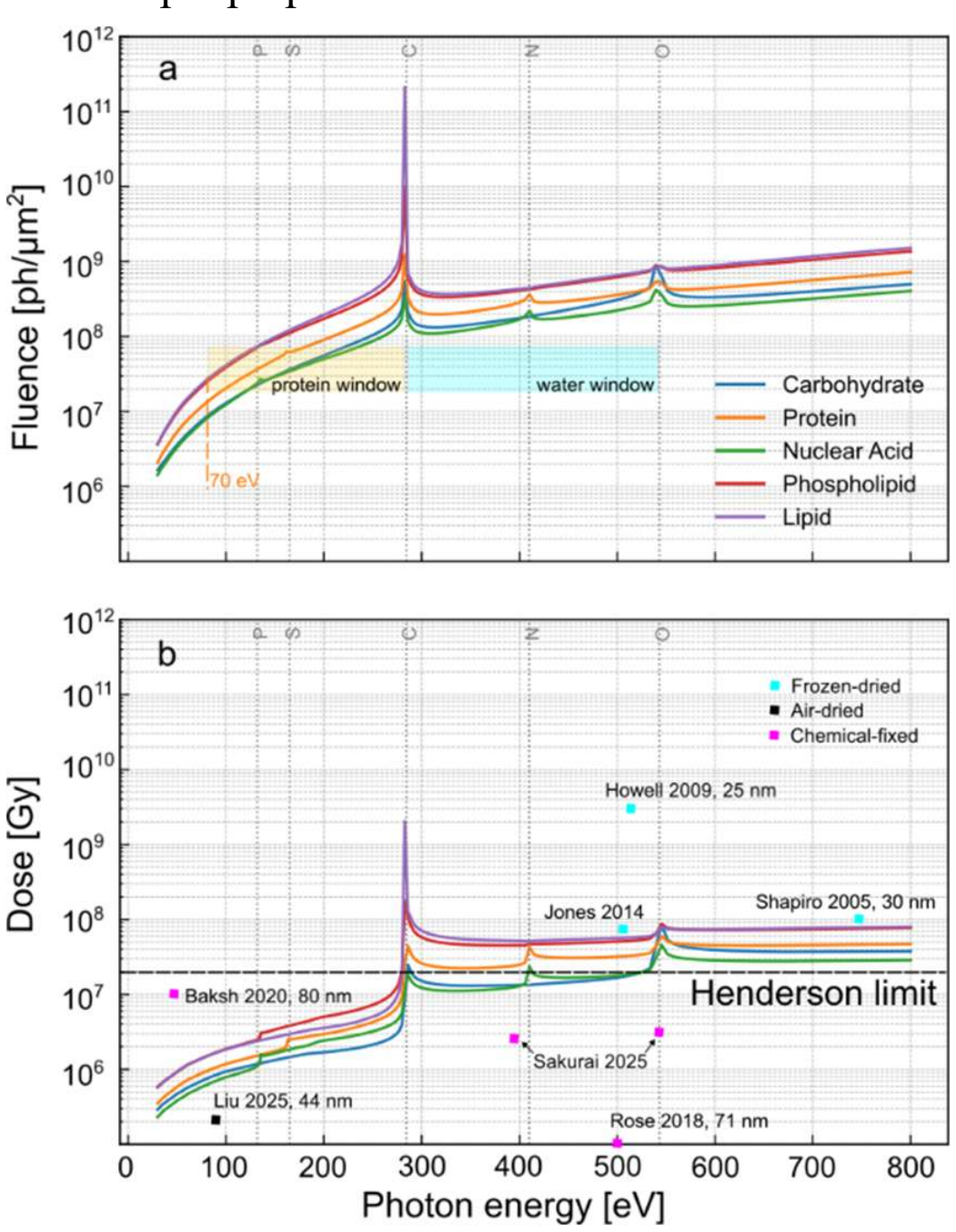

Fig. 7 Fluence (a) and dose (b) requirements for resolving 10 nm half-pitch biological molecules in dehydrated cytosolic background under Rose criterion (K = 5). Henderson limit (black dashed line) is marked as general damage threshold. Reported dose values from XDM dehydrated cell imaging experiments with claimed resolution (if appliable) are plotted in (b).

## 4. Discussion

Our work compares the contrast and dose requirement in hydrated and dehydrated conditions, highlighting a major benefit of dehydrated state imaging in a practical window (70 – 284 eV), i.e. the EUV spectral regime. With water removed, the cytosolic background is close to vacuum or in other words a low-density medium, allowing EUV photons to penetrate μm-thick specimens. Consequently, fewer photons are needed to achieve a sufficient image SNR (Rose criterion). Dehydrated specimens therefore permit high-resolution imaging with a significant reduced dose. Moreover, radiation damage is less severe without water, since radiolysis of water and diffusion of radicals are major damage pathways in hydrated samples. Dehydrated cells can thus tolerate orders of magnitude higher doses, with negligible mass loss or shrinkage reported even up to $10^7$ Gy [9].

Our study shows that these effects are especially pronounced in the EUV regime, where we claim that 10 nm resolution in dehydrated cells can be achieved with a dose below the Henderson limit, i.e. at room temperature without significant radiation damage. For embedded tissue slices, resolution is more limited but still feasible under relaxed dose constraints, provided the sample is prepared in a suitable medium, e.g., protein-rich tissues embedded in paraffin.

Another key advantage of dehydrated-state imaging in this regime is the enhanced biomolecular contrast. The refractive index and absorption differences between biomolecules and vacuum are much greater than between biomolecules and water, yielding enhanced contrast for subcellular features. Combined with advanced lensless imaging modalities like ptychography [72–74], this improved biomolecular contrast means that staining is unnecessary.

In summary, our work opens up a practical protein window for dehydrated biological imaging, which possesses the advantages of simple preparation, dose efficiency and enhaned contrast between cellular material and background, and most importantly, not limited by strong water absorption.

Building on first demonstrations of tabletop EUV ptychography on microorganisms [38,39], our study provides a theoretical framework that establishes EUV diffraction imaging on dehydrated biological specimens as a potentially compelling and simple pathway for cryo-free imaging with 10 nm resolution. Future studies can extend this label-free, high-resolution technique to more complex biological contexts, i.e., visualizing the interaction between microorganisms and host cells, and capturing nanoscale dynamics induced by drug treatments [45]. Such advances open avenues for clinical translation, where compact EUV lensless imaging systems might complement conventional pathology by enabling label-free diagnostics, i.e., detecting cancer-related architectural changes in thin tissues.

**Acknowledgments**

The work is supported by the Free State of Thuringia and the European Social Fund Plus (2023FGR0053).